\documentclass[twocolumn,aps,prl,amsmath,nofootinbib,showkeys]{revtex4}

\usepackage[utf8]{inputenc}
\usepackage{graphicx}
\usepackage{bm}                      
\usepackage{mathptmx}                
\usepackage{newtxmath}
\usepackage{amssymb}  
\usepackage{dcolumn}                 
\usepackage{color}
\usepackage{comment}

\newcommand{\be}{\beta}

\newcommand{\gsim}{\, \raisebox{-0.8ex}{$\stackrel{\textstyle >}{\sim}$ }}
\newcommand{\lsim}{\, \, \raisebox{-0.8ex}{$\stackrel{\textstyle <}{\sim}$ }}

\newcommand{\beq}{\begin{equation}}
\newcommand{\eeq}{\end{equation}}
\newcommand{\ba}{\begin{array}}
\newcommand{\ea}{\end{array}}
\newcommand{\bea}{\begin{eqnarray}}
\newcommand{\eea}{\end{eqnarray}}
\newcommand{\bi}{\begin{itemize}}  
\newcommand{\ei}{\end{itemize}}
\newcommand{\ben}{\begin{enumerate}} 
\newcommand{\een}{\end{enumerate}}
\newcommand{\bc}{\begin{center}}
\newcommand{\ec}{\end{center}}

\newcommand{\ee}[1]{\times 10^{#1}}

\def\bea{\begin{eqnarray}}
\def\eea{\end{eqnarray}}
\def\be{\begin{equation}}
\def\ee{\end{equation}}

\def\beq{\begin{equation}}
\def\eeq{\end{equation}}
\def\bar{\begin{array}[b]}
\def\barc{\begin{array}}
\def\bart{\begin{array}[t]}
\def\ear{\end{array}}

\begin{document}

\title{Isotropic Lifshitz point in the $O(N)$ Theory}

\author{Dario Zappal\`a}

\affiliation{
INFN, Sezione di Catania, Via Santa Sofia 64, 95123 Catania, Italy}

\date{\today} 

\begin{abstract}
The presence of an isotropic tricritical  Lifshitz point for the $O( N ) $  scalar theory is investigated 
 at large $N$\, in the improved Local Potential Approximation (LPA$'$) 
 by means of the Functional Renormalization Group equations.
At leading order, the non-trivial Lifshitz point is observed if the number of  dimensions 
$d$ is taken between $d=4$ and $d=8$, and the eigenvalue spectrum of the associated 
eigendirections  is derived. 
At order  $1/N$ of the LPA$'$  the anomalous dimension $\eta_N$ is computed and it
is found to vanish both in $d=4$ and $d=8$.
The dependence of our findings on the infrared regulator is discussed.
\end{abstract}

\maketitle

{\bf Introduction} -- 
The description of a tricritical Lifshitz  point  by a Landau-Ginzburg  $\phi^4$ model, where the derivatives of the field with respect to 
the coordinates of a $m$-dimensional subset of a $d$-dimensional space and those of the complementary  
$(d -m)$-dimensional  subspace  possess different scaling laws, was first presented  in \cite{Horn}. 
More specifically, in \cite{Horn} the kinetic term with  square gradient of  the field, $O(\partial^2)$, is kept finite only for the second  
subset of coordinates, while the corresponding term of the $m$-dimensional subset is suppressed, so that the term with four powers  
of the gradient,  $O(\partial^4)$, becomes the leading kinetic term of the $m$-dimensional subspace and this induces  drastic changes 
in the scaling properties of the theory.

The Lifshitz points,  which are related to the coexistence on the phase diagram of three phases, one with vanishing order parameter,
another  with finite constant order parameter and the third characterized  by a modulated order parameter with finite wave vector,
find application in  various fields such as magnetic systems as well as polymer mixtures or high $T_C$ superconductors (for reviews see  
\cite{selke1988,Diehl}),  but, recently, also in different contexts such as Lorentz symmetry violation, \cite{Alexandre,kikuchi}, or  emergent 
gravity theories \cite{horava, filippo,zz,bekaert1,cognola}.  In addition, an oscillating phase has been predicted for a very wide class of systems 
\cite{casal1,mizu,castorina,casal2,buballa}, and it is conceivable to  expect  that  a Lifshitz point could be associated to these modulated phases. 
In this sense, a more complete understanding of the properties of the Lifshitz point is certainly desirable.

Rather than considering the general case with $0<m<d$, where the different scaling properties in the two separate subspaces lead to 
a peculiar critical behaviour that involves two different anomalous dimensions and correlation lengths, we shall focus on  the isotropic  
case  with $m=d$. In fact, if $m<d$, due to the different behaviour of the two sets of coordinates, the isotropy of the problem is lost while, 
when $m=d$, all the space coordinates have the same critical behaviour  and spatial isotropy  is preserved.  Clearly, in this latter case the 
critical scaling remains  different from the standard one because, as explained before,  the kinetic term in the action is quartic, rather than 
quadratic,  in the field  derivatives.

The critical properties of the Lifshitz point were studied in the $\varepsilon$-expansion \cite{Horn} as well as in the $O(1/N)$ expansion \cite{diehl4}. 
The isotropic case $m=d$ was considered within an expansion around $d=8$ and  $\varepsilon=8-d$ \cite{Diehl3}  while, recently, a numerical 
Monte-Carlo study indicated a possible disappearance of the Lifshitz point, when fluctuations are properly included \cite{schmid}. 

Furthermore, another non-perturbative technique already employed to study this problem is  the Functional Renormalization Group (FRG) 
\cite{Wetterich:1992yh,Morris:1994ie,Berges:2000ew}
which  consists of a  set of differential flow 
equations either for various operators entering the effective 
action of the theory,  or for one or more $n$-point Green functions derived from the effective action. Fixed points correspond to 
stationary points of these equations and the critical exponents, that classify relevant, marginal and irrelevant operators,
are extracted by determining the eigenvalue spectrum of the linear reduction 
of the differential equations around the fixed point solutions.
Coming to the Lifshitz point, the FRG  was  applied to study  this problem for a one component scalar  theory, 
$N=1$, \cite{Bervillier}, and for the  $N=3$ theory, \cite{Essafi}, both in the uniaxial ($m=1$) case.
Finally, the isotropic  case ($m=d$) with  $N=1$ was considered in  \cite{bonzap} and,
in this last case,  the Proper Time version \cite{liao,bohr,boza2001} of the FRG, which can be
formally derived in  the framework of the background field flows\cite{Litpaw3,Litpaw4}, 
was used because it proved to be quite accurate and suitable for the numerical analysis of the critical 
properties of a theory at  a fixed point \cite{boza2001,zap2001,Maza,litimzappala} and, in addition,
the  Proper Time flow equation of the  $O(\partial^4)$ operator 
(coupled to the potential and to the  $O(\partial^2)$ operator equations), 
that is necessary to treat a Lifshitz point,  had been already derived in \cite{litimzappala}.

The numerical analysis performed in \cite{bonzap}  for the $N=1$ theory, shows at the 
lowest order (in the Local Potential Approximation - LPA), i.e. by considering the  fixed potential equation only,
that a non-trivial solution exists  when the number of spatial dimensions  is $4<d<8$, and for 
$d\geq 8$ the solution merges with the trivial, gaussian fixed point, while for $d\leq 4$
the asymptotic structure of the differential equation changes and no discrete set of
non-trivial  solution is available.  Then, when going beyond the LPA and including the 
differential equations  for the $O(\partial^2)$ and  $O(\partial^4)$ operators, a  solution
was observed in the range $5.5<d<8$, but the numerical analysis for smaller $d$ becomes 
too demanding and it was not possible to establish whether the Lifshitz point survives down to 
$d=4$ or, rather, the fluctuations associated with  higher derivatives terms, $O(\partial^2)$ and  $O(\partial^4)$,
effectively destroy the critical behaviour when  $d$ approaches $4$.

In this letter we consider another aspect of the problem and analyze the existence of a  Lifshitz point 
for a scalar $O(N)$-symmetric theory,  in order  to find out whether the critical behaviour survives to 
the presence of the strong infrared fluctuations due to the transverse modes. 
To this aim, a numerical analysis would require the resolution of a very large number of  differential equations 
that would probably present the same kind of problems observed for the simpler $N=1$ case. 

Therefore, we follow a different approach. We start by considering the procedure developed 
in  \cite{blaizot1,blaizot3}, where the flow equation for the effective action is projected  onto a set of flow equations 
for the $n$-point Green functions which is to be truncated at some specific $n$
and, for our purposes, we retain  the three equations for the potential and the longitudinal and
transverse two-point functions. Then, we treat these equations in the framework of the $1/N$ expansion
to extract the corresponding $O(1/N)$ contribution to the anomalous dimension $\eta$.

Actually, this combined procedure  neither amounts to a full resolution of the flow equations 
for the two point functions derived in \cite{blaizot1,blaizot3}, nor to a complete $O(1/N)$ computation,
however it allows us to go one step beyond the leading order of the $1/N$ expansion (at which the full
eigenvalue  spectrum of the Lifshitz point is determined) and establish both the survival of the 
Lifshitz point and the first non-vanishing contribution to $\eta$ at criticality.
In particular we find that the expression derived for $\eta$ according to this procedure reduces to 
the result that is obtained in the minimal improvement of the LPA \cite {wettapp1}, also known as LPA$'$.

\vskip 5 pt

{\bf Flow equations} -- 
In order to write down the fixed point equation, 
we start from the full FRG  equation 
\cite{Berges:2000ew}:
($ \partial_t\equiv k \partial_k$):
\beq \label{rgfloweq}
\partial_t \Gamma_k[\phi]=\frac{1}{2} \int_q\, \partial_t R_k(q)
\left [ \Gamma_k^{(2)}[q,-q;\phi]+R_k(q)\right ]^{-1}
\eeq
$\Gamma_k[\phi]$ being the running  effective action at scale $k$,
and  $R_k(q)$  a suitable regulator that suppress the
modes with $q\ll  k$ and allows to integrate those with  $q\gg k$.
The specific choice of the regulator  $R_k(q)$ is discussed below.

Rather than introducing the running parameters by means of an explicit form of the effective action,
we proceed by  displaying the second functional derivative of the effective action
$ \Gamma_{ab}^{(2)} (p;\phi)\equiv  \delta^2\Gamma_k / \left ( \delta\phi_a (p)
\delta\phi_b(-p) \right )$ that, according to the $O(N)$ symmetry, has the general form 
($\rho\equiv \phi_a \phi_a/2$) :

\beq\label{gamma2n}
\Gamma_{ab}^{(2)} (p,\phi) = \Gamma_{A} (p,\rho)\delta_{ab} +
\phi_a \phi_b \Gamma_B (p, \rho)
\eeq

Then, we parametrize $\Gamma_{A}$ and $\Gamma_{B}$ in terms of the
potential $V$ and of the renormalization functions $Z_A,\, Z_B,\,W_A,\,W_B$, i.e. the coefficients of the quadratic and quartic  powers of the momentum $p$:
\beq\label{gammaab1}
\Gamma_A(p^2, \rho) =  W_A(\rho) \,p^4 + Z_A( \rho) p^2 +   V'  
\eeq
\beq\label{gammaab2}
\Gamma_B(p^2, \rho) =  N\, W_B(\rho) \,p^4 + N \, Z_B( \rho) p^2 + V''
\eeq
where prime indicates the derivative with respect to $\rho$ and $N$ is the number of field components.

The factor $N$ appearing in front of $W_B$ and $Z_B$ is due to a specific rescaling of the potential 
and of the field with respect to the standard definitions,  $V\to N\,V$ and $\phi_a \to \sqrt{N} \phi_a$, which 
is made  in order to derive a fixed point equation that is directly arranged in a $1/N$ expanded structure.  
Clearly, this rescaling has no effect on $V'$, while it changes $V'' \to V''/N$ as well as the factor 
$\phi_a \phi_b \to N\, \phi_a \phi_b $ in Eq.\,(\ref{gamma2n}), these last two transformations being responsible 
for  the factor $N$ appearing in the definition of $\Gamma_B$  in Eq. \,(\ref{gammaab2}). 
Therefore,  from Eqs. \,(\ref{gammaab1}) and (\ref{gammaab2}), it  is easy to expect the parameters 
$W_B, \, Z_B$ to be  $1/N$ suppressed with respect to $W_A, \, Z_A$, as it will be checked below.

Then, if we separate the longitudinal (L) and transverse (T) components in the 
inverse of $\Gamma_{ab}^{(2)} (p,\phi)$, that is the propagator of the theory $G_{ab}(p,\phi)$,  according to:
\beq\label{propagatoren}
G_{ab}(p,\phi)=\Big(\delta_{ab}  -\frac{\phi_a\phi_b}{2\rho} \Big)  \,G_T(p,\rho)
+ \frac{\phi_a\phi_b}{2\rho} \,G_L(p,\rho)\eeq
one finds
\beq\label{gt}
G_T^{-1}(p,\rho) = \Gamma_A(p,\rho)
\eeq
and 
\beq
\label{gl}
 G_L^{-1}(p,\rho)= \Gamma_A(p,\rho) + 2\rho \Gamma_B(p,\rho) \; .
\eeq
It is understood that the field dependent parameters $V, \,W_A, \, W_B, \, Z_A, \, Z_B$ 
also depend on the running scale $k$ and, with these settings, we can rely on the derivation of the flow equations 
carried out  in \cite{blaizot3}. 
We define the integrals 
\beq
\label{intj}
J_n^{\alpha \beta}(p,\rho)=
 \int_q \partial_t R_k(q) \widetilde G_{\alpha}^{n-1}(q,\rho) \widetilde G_{\beta}(p+q,\rho) \; ,
 \eeq
\beq\label{inti}
I_n^{\alpha \beta}(\rho)=J_n^{\alpha \beta}(0,\rho),
\eeq
where  $n\geq 1$,
$\int_q\equiv\int \frac{d^dq}{(2\pi)^d}$, 
$\alpha$ and $\beta$  stand either for $L$ or $T$, and 
\be\label{inverse}
(\widetilde G_{\alpha}^{n}(q,\rho)\,)^{-1}\equiv
(G_{\alpha}^{n}(q,\rho)\,)^{-1} + R_k(q) \;\; .
\ee

Then, by following \cite{blaizot3} (see also \cite{zappa2012}), we get the flow equation for the potential $V$:
\begin{equation}
\label{o4vflow}
\partial_t V(\rho) = \frac{1}{2}\left \{ I_1^{TT}(\rho) + \frac{1}{N} \left [ I_1^{LL}(\rho) - I_1^{TT}(\rho)  \right]  \right \}
\end{equation}
and for the two-point functions, properly subtracted of the zero-momentum contribution:
\beq
\label{gammaflow}
 \partial_t \left[ \Gamma_X(p^2, \rho)  - \Gamma_X(0, \rho) \right] =F_X(p^2, \rho)  - F_X(0, \rho) 
\eeq
where $X$ stands either for $A $ or $B$, and
\bea
\label{zaflow}
&  \displaystyle    
F_A(p^2, \rho)=
 - \frac{1}{2}I_2^{TT} \,  \Gamma_A'     
 +\frac{1}{N}  \Bigg [
2  \rho \left ( J_3^{LT} \, {\Gamma_A'}^2 + J_3^{TL}\, {\Gamma_B}^2  \right ) 
 \nonumber\\
&   \displaystyle  
-I_2^{LL} \left( \frac{ \Gamma_A'}{2} + \rho  \Gamma_A''  \right) -
I_2^{TT} \left ( \Gamma_B  -\frac{\Gamma_A'}{2}  \right )   \Bigg ]  \;,
\eea
\beq\label{zbflow}
 F_B(p^2, \rho) = 
J_3^{TT}\, {\Gamma_B}^2 
- \frac{1}{2} I_2^{TT} \,\Gamma_B' + O\left (\frac{1}{N}\right)
\eeq
Eqs.\,(\ref{gammaflow}), (\ref{zaflow}), (\ref{zbflow}) can be 
reduced to flow equations either for $W_X$ or  $Z_X$, by selecting in $F_X$ 
the terms proportional respectively to $p^4$ or $p^2$. Then,
it is evident from  Eqs. \,(\ref{gammaab2}), (\ref{gammaflow}) and  (\ref{zbflow})  that,
in order to avoid any inconsistency in the $1/N$ expansion, 
$W_B$ and $Z_B$ must be  $O(1/N)$  so that  $\Gamma_B\sim O(1)$.
Accordingly, we are allowed to neglect $O(1/N )$  corrections in Eq. (\ref{zbflow}), as they 
contribute to $W_B$ and $Z_B$ to order $O(1/N^2)$.

Let us now consider the regulator $R_k(q)$. A particularly useful regulator,
that has the advantage of reducing the integrals to simple structures which 
can be analytically solved in most cases, was introduced in  
\cite{litim2}
and  has the form:
\beq
\label{optimctf}
R^\theta_k(q) = (k^2 -q^2)\, \, \widehat Z_k \,\,\theta (k^2 -q^2)
\eeq
where $\theta$ is the Heaviside step function
and a  $k$-dependent (but field independent) 
normalization factor  $\widehat Z_k$ is included.
For the present problem the regulator  in Eq.\,(\ref{optimctf})
should be modified into 
$R^\theta_k(q) = (k^4 -q^4)\, \,
\widehat W_k \,\,
\theta (k^2 -q^2)$
with $\widehat W_k$   taken   equal to $W_A$, evaluated at a particular value of $\rho$ :
$\widehat W_k =W_A(\overline \rho)$, with  $\overline \rho$ to be specified.
However, due to the presence of the Heaviside function, 
the second and higher derivatives of 
$R^\theta_k(q)$ with respect to $q^4$,
generate a singular behaviour of the integrals involved in 
this analysis. Therefore it is preferable to replace $R^\theta_k(q)$
with a smooth, one-parameter ($\alpha$) regulator:
\beq
\label{ourctf}
R_k(q) =\frac{\widehat W_k }{2} \left [ (k^4 -q^4) 
+\sqrt{(k^4 -q^4)^2 + (2\alpha)^{-2} } \right]
\eeq
In fact, $R_k(q)$ in Eq.\,(\ref{ourctf})  approaches 
$R^\theta_k(q)$ in the limit $\alpha^{-1}\to 0$,
and, for values of the dimensionless parameter $k^4 \alpha\sim 10^3$ or larger,
$R_k(q)$ (and its first derivative) can be practically replaced  by $R^\theta_k(q)$ 
in the resolution of the integrals but,
on the other hand, all its derivatives are regular so that it does not generate any singularity as long 
as $\alpha$ is kept finite, i.e. $\alpha^{-1} \neq 0$.

Incidentally, as the vanishing of of the regulator at $k=0$, $R_{k=0}(q)=0$, is a necessary requirement of the flow equations, 
then  $\alpha^{-1}$ must be  a  function of the scale $k$ that vanishes at $k=0$. This can be easily  achieved e.g. by taking  
$(2 \alpha)^{-1}= \lambda\Lambda^4 \, {\rm tanh}[(k /\Lambda)^\mu]$, where $\Lambda$ is a fixed mass scale and $\lambda$ 
and $\mu$  two small dimensionless parameters that can be adjusted to set the size of $(2 \alpha)^{-1}$  and of  its derivatives. 
Due to the dependence of $\alpha$ on $k$,  the fixed point equations  do contain additional terms proportional to  
$k \, [ \, \partial (2 \alpha)^{-1} /\partial k\, ] =  2 \, \mu \,( k/\Lambda)^\mu \, (2 \alpha)^{-1} \,{\rm sinh}^{-1} [2\,(k /\Lambda)^\mu]$, 
but one easily realizes  that even the largest contributions  (proportional to $\alpha$)  encountered in the following calculations, 
when multiplied by this factor, for sufficiently small values of $\mu$ turn out to be systematically suppressed with respect to the 
other terms  appearing in the  the fixed point equations. Therefore, we neglect the contributions proportional to 
$k \, [ \, \partial (2 \alpha)^{-1} /\partial k\, ]$ and simply treat  $\alpha$ as a free parameter.

However, as discussed below,  even the regulator in Eq.\,(\ref{ourctf})  is not sufficient to get rid 
of all potentially large (divergent in the limit $\alpha^{-1} \to 0$) terms and therefore at some point 
we find convenient to analyze our equations by adopting the smoother exponential regulator 
\beq
\label{expctf}
R^b_k(q) = \frac{b\,  \widehat W_k q^4}{e^{q^4/k^4}-1}
\eeq
where $b$ is a dimensionless adjustable parameter.

\vskip 5pt

{\bf Leading order of the $1/N$ expansion} -- As anticipated, Eqs.\,(\ref{o4vflow}), (\ref{gammaflow}), (\ref{zaflow}) and  (\ref{zbflow}), 
are already arranged in a $1/N$ expansion structure and we can straightforwardly extract the leading 
($1/N=0$) flow equations for the suitably rescaled parameters,  and  also the associated
fixed point equations, which are obtained by requiring the
rescaled parameters to be  $t$-independent.
The rescaled parameters, relevant for our analysis, are $\varrho=k^{-d+4-\eta}\rho$,
$v=k^{-d}\,V$, $w^A=k^{\eta}\,W_A$, $w^B=k^{d-4+2\eta}\,W_B$, 
$z^A=k^{\eta-2}\,Z_A$, $z^B=k^{d-6+2\eta}\,Z_B$, 
where the scaling dimensions, i.e.  the exponents in the  powers of the scale $k$, are given in 
\cite{Essafi,bonzap},  and the fixed point equations for $w^A$   and  $z^A$ at  $1/N=0$  are:
\begin{equation}
\label{fpwa}
-\eta_0 w^A_0 + (d-4 +\eta_0)\varrho {w^A_0}'=- \frac{1}{2}I_2^{TT} \, {w^A_0}'
\end{equation}
\begin{equation}
\label{fpza}
(2-\eta_0) z^A_0 + (d-4 +\eta_0)\varrho {z^A_0}'=- \frac{1}{2}I_2^{TT} \, {z^A_0}' \;\; .
\end{equation}

In Eqs. (\ref{fpwa}) and  (\ref{fpza})  the prime indicates  derivation
with respect to $\varrho$  and the  subscript $0$ indicates the lowest order of the $1/N$ expansion.
It is easy to check that a field independent $w^A_0$ (and therefore  ${w^A_0}' =0$) together with $\eta_0=z^A_0=0$
is a solution of this set of equations. Therefore, we can take  $w^A_0=1$ to set the overall normalization of the effective action.

Then, we turn to the fixed point equation for the potential, Eq.\,(\ref{o4vflow}), and, after setting 
$\widehat W_k=1$ in  Eq.\,(\ref{ourctf}),  the integral $I_1^{TT} $ can be solved and 
Eq.\,(\ref{o4vflow}) conveniently written as :
\begin{equation}
\label{fpv}
\left [ \left (x+f(x) \right)^2\,d_+ - 1\right ] \, f(x) = \left [ (x+f(x) )^2\,d_- - 1\right ] \, x f_x(x)
\end{equation}
with the following definitions $x=\sqrt{2\varrho}$; $f(x)={\rm d} v / {\rm d} x$; $f_x(x)={\rm d} f / {\rm d} x$; $d_\pm=(d\pm4)/(2\tau)$
and finally $\tau=2/[(4\pi)^{d/2} \Gamma(1+d/2)]$ is the factor coming from the resolution of the integral $I_1^{TT} $.

Eq.\,(\ref{fpv}) can be easily attacked numerically, but all the essential features can be deduced by simple inspection.
In fact we immediately see that the constant function  $f_G(x)=0$ is a solution of Eq.\,(\ref{fpv}), that plays the same 
role of the gaussian fixed point for the standard scaling.
In addition, we observe that a viable non-trivial Lifshitz solution  $f_L(x)$  must vanish at the origin $f_L(0)=0$ 
 due to the symmetry of the problem and, in addition, another  zero of $f_L$  must occur at 
\be\label {xbar}
\overline x^2=\frac{2\tau}{(d-4)}
 \ee
i.e. $ f_L(\overline x)=0$ with non-vanishing derivative $ {f_L}_x(\overline x)\neq 0$. 
By expanding  Eq.\,(\ref{fpv}) around $\overline x$, one finds from the linear terms: 
\be
\label{fprimo}
{f_L}_x(\overline x)=\frac{8-d}{d-4} \;.
\ee

Eq.\,(\ref{xbar}) loses meaning  when $d\leq 4$, while the vanishing of ${f_L}_x(\overline x)$
from  Eq. (\ref{fprimo}) at $d=8$  indicates a flattening of the solution ${f_L}$ 
onto the trivial solution $f_G$. The latter result accords with the numerical analysis of \cite{bonzap}
with $N=1$,  which indicates that the two solutions merge at $d=8$ and  
only the trivial solution survives for $d\geq 8$. Therefore we limit the study 
of Eq.\,(\ref{fpv})  to the range $4\leq d\leq 8$.

With the information collected above, we are able to determine the eigenvalues $\lambda_L$ 
of the flow equation, linearized around the fixed point solution.
To this aim we follow the procedure originally worked out in  \cite{morrisdatt,morristurner} for the standard 
Wilson-Fisher (WF) fixed point, (see also \cite{Maza}) and, 
by writing the $t$-dependent function 
$f(t,x)=f_L(x)+e^{\lambda t}h(x)$,
 as the  sum of the fixed point solution $f_L(x)$ and a perturbation $h(x)$, 
 we  get the the following linear (in $h(x)$ )  equation: 
\bea
\label{lf}
&{\displaystyle \frac{\lambda \,h}{\tau} }
=\left [d_+  -  \left ( x+f_L \right )^{-2} \right ]\,h -
\left [d_-  -  \left ( x+f_L \right )^{-2} \right ]\,x\,h_x
\nonumber\\
&+2 \left ( x+f_L \right )^{-3} \, \left ( f_L \, h - x {f_L}_x \, h \right ) 
\eea
The function $h$ is supposed to be regular  at any finite $x$ and 
can be expanded around $\overline x$: 
\be
\label{acca}
h(x)=\sum_{i=n}^{\infty}a_i (x-\overline x)^i
\ee
where the  lowest power $n$ must be a non-negative integer, $n \geq 0$.
At $x=\overline x$ the coefficient of $x \,h_x$ in square brackets vanishes
and therefore $h_x (x)/h(x)$ 
is either singular  at $x=\overline x$
(with a simple pole singularity) 
or  finite, the former case corresponding to $n>0$ and the latter to 
$n=0$ in Eq.\,(\ref{acca}). In both cases,
after dividing  both members of Eq.\,(\ref{lf}) by $h$,
one can make the replacement  $h_x (x)/h(x)= n/(x-\overline x)$ 
in order to compute the linear corrections in the expansion of Eq.\,(\ref{lf})
around the point $\overline x$. This expansion, with the help of 
Eq.\,(\ref{fprimo}),  yields the following  eigenvalue spectrum (we recall
$4\leq d\leq 8$ ) :
\be
\label{eigen}
\lambda_L=d-4-4n
\ee
parameterized by the non-negative integer $n\geq 0$. 
By following the same procedure, one derives
from Eq.\,(\ref{lf}) the eigenvalues associated  
to  $f_G$ (again with integer $n\geq 0$) :
\be
\label{eigeng}
\lambda_G=4-(d-4)	\,n
\ee
In particular one can determine  those values of $n$ that correspond to relevant (positive) eigenvalues,
namely  $0\leq n<(d-4)/4$ from Eq.\,(\ref{eigen}), and  $0\leq n<4/(d-4)$  from Eq.\,(\ref{eigeng}).
In addition, we  observe that in $d=8$ the two spectra 
in  Eq.\,(\ref{eigen}) and (\ref{eigeng}) are equal, as the two 
fixed point solutions become coincident.

In conclusion, the solutions found  at $1/N=0$
with this particular scaling, clearly resemble those obtained 
with standard scaling where, aside from the constant gaussian solution 
with eigenvalue spectrum $ \lambda_g=2-(d-2)	\,n$, one
has
the WF fixed point  with $\lambda_{WF}=d-2-2n$.
One clearly sees that the difference, at this order, is only  
in the range spanned by $d$  which, in this case, goes from $d=2$ to $d=4$,
while, in the  analysis of the tricritical Lifshitz  point,  from $d=4$ to $d=8$.
In fact, even the number of relevant directions is the same in the two cases,
once the proper change in $d$ is taken into account.

 \vskip 5 pt
 
{\bf $1/N$ corrections}  -- At the leading order $1/N=0$, the equations for the momentum dependent parts
admit the elementary field-independent solutions 
$w^A_0=1$ and $z^A_0=0$, together with $\eta_0=0$, while the equations for  
$w^B$ and $z^B$ at this order decouple from the other equations and one is left 
with the fixed point equation for the potential only.

For the next step, we consider the potential expansion $v=v_0 +  v_N /N+ O(1/N^2)$
and the analogous expansions for $\eta , \, w^A ,\, z^A ,\, w^B ,\, z^B$,
and insert them into  the fixed point equations in order to analyze the  $1/N$ corrections.
We start by observing that 
Eq.\,(\ref{o4vflow}) for $v_N$ involves
the  $(1/N)$ corrections of all the above variables (we recall here that
$w^B_0= z^B_0=0$, but  the 
first non-vanishing  terms of the expansion of $w^B$ and $z^B$, which are  $O(1/N)$, 
contribute to the  leading order  ($1/N=0$)
longitudinal propagator $G_L$,  because of the factor $N$ in Eq.\,(\ref{gammaab2})).
Therefore, a full determination of the $1/N$
corrections requires the resolution of 
five coupled equations.

However,  it is possible  to 
determine $\eta_N$ without solving the whole set of equations.
To this aim,
a direct inspection of equations
(\ref{gammaflow}), (\ref{zaflow}), (\ref{zbflow}) 
shows that  the vanishing field-independent solution
$w^A_N = z^A_N = w^B_N = z^B_N=0$ is  not allowed
because of the non-vanishing coefficients of the integrals  $J_3$ in Eqs.\,(\ref{zaflow}), (\ref{zbflow}), 
respectively ${\Gamma_A'}^2$ and ${\Gamma_B}^2$, which are finite and field dependent 
due to their dependence on $V'$ and $V''$, as shown in  Eqs.\,(\ref{gammaab1}), (\ref{gammaab2}).

Nevertheless, at least  for  one particular value of the field   $\varrho=\overline \varrho$,
we can extend at $1/N$ the normalization of the propagators, already fixed by the 
leading order solution $w^A_0=1; \,z^A_0=w^B_0= z^B_0=0$. This immediately  implies 
$w^A_N(\overline \varrho )= z^A_N(\overline \varrho )= 
w^B_N(\overline \varrho )= z^B_N(\overline \varrho )=0$
and it is natural to take  $\overline \varrho$ as the  point  where the  derivative of the leading  order 
potential vanishes, i.e. $\overline \varrho= \overline x^2 /2$,  with $\overline x$ defined in Eq.\,(\ref{xbar}).
Finally, we  extract from Eq.\,(\ref{zaflow}) the two  equations for 
$w^A_N,\, z^A_N$, directly computed at $\overline \varrho$ :
\bea
 \label{etan}
&-\eta_N - \,(d-4) \,\overline \varrho \,{w^A_N}' (\overline\varrho)= 
\nonumber\\
&
- \frac{1}{2}(I_2^{TT})_0 \,\, {w^A_N}' (\overline\varrho)
+ 2  \,\overline \varrho \, v_0''(\overline\varrho)^2 \,\left ( J_3^{LT}\big |_{p^4}  + J_3^{TL}\big |_{p^4}  \right )_0 
\eea

 \bea
 \label{zetan}
&- \,(d-4) \,\overline \varrho \,{z^A_N}' (\overline\varrho)= 
\nonumber\\
&
- \frac{1}{2}(I_2^{TT})_0 \,\, {z^A_N}' (\overline\varrho)
+ 2  \,\overline \varrho \, v_0''(\overline\varrho)^2 \,\left ( J_3^{LT}\big |_{p^2}  + J_3^{TL}\big |_{p^2}  \right )_0
\eea
where the subscript $0$ of the various  integrals indicates that they must be computed by using the leading 
order ($1/N=0$) solution of the various parameters, while the subscript $p^4$  in Eq.\,(\ref{etan})
and $p^2$ in Eq.\,(\ref{zetan}) of the integrals $J_3$, 
indicates that only the  the coefficient of that particular  power of the momentum $p$ 
in the expansion of the addressed  integral is to be retained.
We find that the $1/N$ correction to the anomalous dimension $\eta_N$
does not appear in Eq.\,(\ref{zetan}), but it is directly obtained from 
Eq.\,(\ref{etan}),  if one neglects  the terms proportional to 
${w^A_N}' (\overline\varrho)$.  
At the same time, the $O(1/N)$ corrections to the fixed point potential in 
Eq.\,(\ref{o4vflow}) are under control and  easily computable  by numerical integration. 

As anticipated, we notice that the procedure adopted to compute  $\eta_N$
essentially coincides with the scheme introduced in \cite {wettapp1} which leads to 
the  improved Local Potential Approximation LPA$'$.
This can be straightforwardly checked by replacing Eq.\,(\ref{etan}) with the equation 
obtained by repeating  the previous steps for the case of the anomalous dimension $\eta_{WF}$ at the WF fixed point,
which gives $\eta_{WF} =  - 2  \,\overline \varrho \, v_0''(\overline\varrho)^2 \, ( J_3^{LT}\big |_{p^2}  + J_3^{TL}\big |_{p^2}  )_0 $.
In this case the expansion is to be taken  to order $p^2$ and
the regulator in\,(\ref{optimctf}) can be safely chosen, because it  does not generate any singularity.
The corresponding  integrals can be analytically computed, as shown in \cite {wettapp1},  and one finds 
$( J_3^{LT}\big |_{p^2}  + J_3^{TL}\big |_{p^2}  )_0
=-\tau / (1+ 2  \,\overline \varrho \, v_0''(\overline\varrho))^2$
and, therefore,   $\eta_{WF} =  2 \tau \,\overline \varrho \, v_0''(\overline\varrho)^2  
/ (1+ 2  \,\overline \varrho \, v_0''(\overline\varrho))^2$.
This is exactly the expression of  the anomalous  dimension which is used in the  LPA$'$
\cite{wettapp1,Berges:2000ew,codello1}.

In order to test  the reliability of this procedure,
we can go one step further and replace in $\eta_{WF}$, the particular value of $\overline \varrho$ 
and $v_0''(\overline\varrho))$ that are obtained from the leading order analysis ($1/N=0$) for  the WF fixed point.
Then, instead of  Eq.\,(\ref{xbar}), one has $\overline x^2={\tau}/{(d-2)}$  and  Eq.\,(\ref{fprimo}) becomes 
${f_{WF}}_x(\overline x)=  2  \,\overline \varrho \, v_0''(\overline\varrho)=   {(4-d)}/{(d-2)} $ 
(these changes are due to the different scaling of the various quantities in the two cases 
and also to the different dimension of the  regulator $R_k$ that, 
when derived with respect to the scale $k$, $\partial_t R_k$,
produces a different factor).  Thus, one finds the following $1/N$ correction to the anomalous dimension at the WF
fixed point:
\be
\label{etawf}
\eta_{WF} = \frac{(d-2)(4-d)^2}{4}
\ee
that is to be compared to the full result directly obtained in the $1/N$ expansion, \cite{zinn} ($\epsilon\equiv 4-d$ and $\Gamma$ indicates the Gamma function) :
\be
\label{etazj}
\eta = \frac{4\epsilon}{(4-\epsilon)\,\pi} \;\,\frac{{\rm sin}\,(\pi\epsilon/2) \,\,\Gamma(2-\epsilon) }{\Gamma(1-\epsilon/2)\,\,\Gamma(2-\epsilon/2)}\;\; .
\ee
Remarkably, Eqs.\, (\ref{etawf}) and (\ref{etazj}) have the same behaviour both for $d=2+\delta$ (with $\delta\gsim0$),  i.e. $\eta_{WF}=\eta=\delta$, 
and  for $d\lsim 4$ (with $\epsilon \gsim0$),
i.e. $\eta_{WF}=\eta=\epsilon^2/2$. Instead, in  $d=3$, where the difference 
between Eq.\, (\ref{etawf}) and Eq.\, (\ref{etazj}) is largest, one finds
$\eta_{WF}=1/4$ and $\eta=8/(3\pi^2)\simeq 1/(3.7)$. 
We take  this small discrepancy as the  measure of the reliability 
of the LPA$'$ here considered even in the case of the Lifshitz critical behaviour.

Going back to the Lifshitz fixed point problem, we have to compute 
$\eta_N$ from Eq.\,(\ref{etan})  by  neglecting  the terms proportional to 
${w^A_N}' (\overline\varrho)$.  However, as anticipated, this time
a strong dependence on the regulator is observed. In particular, the parameter $\alpha$ introduced in  Eq.\,(\ref{ourctf})
explicitly shows up in  the resolution of the integrals, because 
$(\partial^2 R_k(q^4)/\partial q^8)_{q^4=k^4}=\alpha$.
Namely, we get
\be
\label{etafin}
\eta_N = \frac{4 \tau \,\overline \varrho \, v_0''(\overline\varrho)^2  }
{D^2}\, 
\left [
4\overline \alpha - \frac{24\,\overline \alpha + d+8}{(d+2)\,  D} +
\frac{6}{(d+2)\, D^2}   
\right ]
\ee
where we introduced the dimensionless parameter $\overline\alpha=k^4 \alpha$
and  $D=(1+ 2  \,\overline \varrho \, v_0''(\overline\varrho))$.
Then, 
with the help of  Eqs.\,(\ref{xbar})  and (\ref{fprimo})
one gets the analogous of Eq.\,(\ref{etawf}) for the Lifshitz case,
with no need to solve the fixed point equation for the $O(1/N)$ corrections to the potential, 
 or the  wave function renormalizations :
\bea
\label{etafin2}
&\eta_N& =
\frac{(d-4)\,(8-d)^2}{16} \, \Bigg \{\,
4\overline \alpha -
\nonumber\\
&& \frac{(d-4)}{4(d+2)} \left ( \, 24\,\overline \alpha+d+8 \,\right )+
\frac{3 (d-4)^2 }{8(d+2)}
\Bigg \}
\eea

We observe the explicit dependence on the parameter $\overline \alpha$ in Eq.\,(\ref{etafin2})
and it is evident that the alternative use of the  Heaviside cutoff $R^\theta_k$, associated to the limit 
$1/\overline \alpha\to 0$ would produce a singular behaviour of $\eta_N$. 
Instead, for finite values of $\overline \alpha$,  $\eta_N$ is finite  in the whole range $4<d<8$.

However the  $\overline\alpha$-dependence in Eq.\,(\ref{etafin2}) 
has strong drawbacks : for instance in $d=6$, one can take $\overline\alpha$
sufficiently large that the term $\eta_N/N$ in  the  $1/N$ expansion of the anomalous 
dimension  is so big, even with $N>>1$,  that the expansion itself become questionable.
The only two cases in which the $\overline\alpha$-dependence  becomes irrelevant are 
the two limits of $\eta_N$ for $d\to 4^+$ and for  $d\to 8^-$, that vanish for 
any fixed value of $\overline\alpha$,  due to the factor in front of the curly bracket in the right 
hand side of  Eq.\,(\ref{etafin2}) .

\begin{table}[htb]
\begin{center}
\begin{tabular}{c c c || c c c}\hline
& {$d=4.1$} & & &$d=5$&\\
\hline\hline\hline
$b$ & & $\eta_N$ & $b$&&$\eta_N$\\
\hline\hline
$10^{-4}$ && $0.081$& $10^{-4}$ && $0.137$\\
\hline
$10^{-3}$ && $0.086$& $10^{-3}$ && $0.230$\\
\hline
$10^{-2}$ & & $0.089$& $10^{-2}$ && $0.313$\\
\hline
$1.5\; 10^{-2}$ && *$0.0894$&$1.2\; 10^{-2}$ && *$0.3144$\\
\hline
$10^{-1}$ && $0.085$& $10^{-1}$ && $0.147$\\
\hline
$2\;10^{-1}$ && $0.080$& $2\;10^{-1}$ && $0.023$\\
\hline
$20$ & & $-0.036$& $20$ && $-0.720$\\
\hline\hline
\end{tabular}
\end{center}
\caption{$\eta_N$ as obtained for different values of $b$ in Eq.\,(\ref{expctf}). }
\label{tab1}
\end{table}  

In order to to collect further indications on the effect of the regulator in the computation of $\eta_N$,
we solve  the integrals  in Eq.\,(\ref{etan})  with the  exponential regulator defined in 
Eq.\,(\ref{expctf}), which has the advantage of being essentially smoother than the one 
in  Eq.\,(\ref{ourctf}) and free of additional dimensionful  parameters but,
on the other hand, no analytical expression for $\eta_N$ can be derived.

Therefore, we report in Table \ref{tab1} the values  of $\eta_N$  obtained 
with different values of the parameter $b$ of the regulator in  Eq.\,(\ref{expctf}) 
for two values of the dimension $d$, namely $d=4.1$ and $d=5$. In both cases 
$\eta_N$ shows the same qualitative behaviour, by reaching a maximum value 
(indicated  by a star)  around $b\simeq 10^{-2}$,  and then systematically 
decreasing down to large negative values.  Unlike the result in 
Eq.\,(\ref{etafin2}) that shows a linear dependence on  $\overline\alpha$,
in this case we can invoke the  minimal sensitivity criterion to select
 the  maximal values as estimates of the anomalous dimension $\eta_N$.

However when $d$ is increased to 6 or to larger values, a more cumbersome picture shows up.
In fact, already at $d=6$  the simple $b$-dependence of Table \ref{tab1}
is lost and one finds three different extrema in $\eta_N$ when $b$ grows 
(two maxima with a minimum in between), namely $\eta_N=  0.154,  -0.282, -0.214,$ 
respectively for $b= 6\, 10^{-3} , 0.41,  2.5$.  We notice that $\eta_N<0$
both at the minimum and at the second maximum and the same pattern of three extrema 
is observed for larger $d$.  

\begin{table}[htb]
\begin{center}
\begin{tabular}{c c c || c c c}\hline
& {$b=0.015$} & & &$b=1$&\\
\hline\hline\hline
$d$ & & $\eta_N/(d-4)$ & $d$&&$-8\,\eta_N/(8-d)^2$\\
\hline\hline
4.1&& $0.8939$&7.9 && $0.9762$\\
\hline
4.05&& $0.9454$& 7.95&& $0.9880$\\
\hline
4.01 & & $0.9888$& 7.99&& $0.9976$\\
\hline
4.005&& $0.9944$&7.995 && $0.9988$\\
\hline
4.001&& $0.9989$& 7.999&& $0.9997$\\
\hline\hline
\end{tabular}
\end{center}
\caption{$\eta_N$ as obtained with $d$ approaching  4 and  8. }
\label{tab2}
\end{table}  

We conclude the analysis with the regulator in  Eq.\,(\ref{expctf}),
by showing in Table \ref{tab2} the behaviour  of $\eta_N$  when $d$ 
approaches the two extremal values $d=4$ and $d=8$.
In the former case, $b$ is obviously selected by the presence of a single maximum 
in $\eta_N$, while in the latter case we took  $b=1$ that corresponds to a rather 
stable (for $d\simeq 8$), negative value of  $\eta_N$, very close to its second maximum
(note that for the remaining two extrema, the effect shown in  Table \ref{tab2} is not observed).
Remarkably, Table \ref{tab2} shows that  the power law behaviour already seen 
in  Eq.\,(\ref{etawf}) for the WF fixed point, and in Eq.\,(\ref {etafin2}) for the 
Lifshitz fixed point with the other regulator, is in fact recovered in this case 
both when $d\to 4^+$  and when $d\to 8^-$.

\vskip 5 pt 

{\bf Discussion} -- 
We investigated the existence of the isotropic tricritical Lifshitz point for the $O(N)$ theory 
in the $1/N$ expansion  and explicitly computed  the associated anomalous dimension in the LPA$'$.
More specifically,  instead of directly implementing  the LPA$'$ to the Lifshitz case,
our analysis started from a set of coupled flow equations for the potential and the two 
point functions derived in  \cite{blaizot1,blaizot3}, which were then evaluated at the next to leading order 
in the $1/N$ expansion and under further assumptions.
This procedure produced an equation for the anomalous dimension $\eta$ that turned out to be equivalent to the 
equation for $\eta$ derived in the LPA$'$. 

It must be remarked that $\eta_N$ determined in the LPA$'$ does not include the full $O(1/N)$ 
corrections to the anomalous dimension and, therefore, an indication of the difference between the 
two determinations was obtained in the case of  the WF fixed point, where the  maximum discrepancy  
amounts to about  8$\%$ at $d=3$ while, close to the extremal values, $d=2$ and $d=4$, 
the two calculations coincide.

We find that, already at leading order, the non-trivial Lifshitz point is observed only between $4<d<8$.
At order  $1/N$ of the LPA$'$,  the presence of the Lifshitz point is confirmed and the anomalous dimension 
$\eta_N$ vanishes both at $d=4$ and $d=8$. This is in agreement with the conjecture that these two 
values respectively represent the lower and  upper critical dimension  for the Lifshitz point of the $O(N)$ theory.

In particular, in  \cite{bonzap}, it is argued for the Lifshitz point of the $N=1$ theory, that the lower critical 
dimension could be associated to the large field behaviour of the fixed potential, corresponding  to the 
particular value of $d$  below which the potential does no longer diverge  as a power law for large values 
of the field $\phi$ but, instead, a continuous set of solutions (constant at large $\phi$) of the fixed potential equation 
is found. This value of $d$ is related to the change of sign of the scaling dimension of $\phi$, that for the case 
considered is $D_\phi=(d-4+\eta)/2$ and therefore, if the anomalous dimension vanishes or is neglected, 
$d=4$ is the requested value.  For the Lifshitz point of the $O(N)$ theory  where, as shown above, 
$\eta_N=0$ in $d=4$, it is  natural to accept it as the lower critical dimension. Needless to say, this
argument is the restatement of what occurs for the scaling of the $O(N)$ theory at the  
lower critical dimension of the WF fixed point, $d=2$.

In addition, one  can focus on the leading order potential equation, Eq.\,(\ref{fpv}), directly in $d=4$. 
Actually,  this equation can be  solved analytically and, as for the case with $d<4$, one ends up with 
a continuous set of  solutions,  parameterized by one real parameter.

Finally we comment on the dependence on the regulator of the result obtained for $\eta_N$.
While we checked that the regulator (\ref{optimctf}) is well behaved for the computation of $\eta_{WF}$ in 
the WF case,  even its smoothened version in Eq.\,(\ref{ourctf}) produces potentially dangerous terms
in the Lifshitz point case; terms that become irrelevant only in the limits $d\to 4^+$ and $d\to 8^-$.
Then, the use of the smoother regulator (\ref{expctf}) on the one hand confirms the 
behaviour of $\eta_N$ in the region close to  $d=4$ and $d=8$  but  on the other hand,  
still produces the undesired effect, at least only  for more than six dimensions, 
of generating multiple spurious extrema in $\eta_N$, regarded as a function of $b$.
Therefore, away from the extremal points $d=4$ and $d=8$,  no firm statement can be made on $\eta_N$
in the LPA$'$.  Conceivably, this is due to the  modified two point functions (\ref{gammaab1}), (\ref{gammaab2}) 
with a leading O($p^4$) term,  which provide the major difference between the Lifshitz point and 
the standard fixed  point case where, conversely, the LPA$'$ provides reliable results.

We conclude by observing that the tricritical Lifshitz point 
which,  when looking at the eigenvalue spectrum at the leading order of the
$1/N$ expansion, could appear  as a trivial duplicate  of the WF fixed point with a 
suitable redefinition of the scaling dimensions of the various operators, does actually 
show original  features.
In fact, not only rather different properties of the anomalous dimension (with respect to the WF case) 
show up at order $1/N$, but it must  also be noticed that,
as soon as the wave function renormalizations are explicitly included in the  fixed point equations,
the coefficient of $(\partial \phi)^2$, $Z$, has  positive scaling dimension $2-\eta$, 
which indicates the existence of a relevant direction that has  no correspondence  at the WF critical point. 
On the other hand the similarities in the two cases could be a hint  that the structure observed  around 
$d=2$,  such as the presence of multi-critical solutions  \cite{morris2dim,codello1,codello2,codello3,defenu},  or the relation 
with phase transitions of different nature \cite{gersdorff,nandori}, could have a counterpart in  the Lifshitz scaling 
around $d=4$.

\vskip 5 pt 

{\it Acknowledgments.}  This work has been carried out within the INFN project QFT-HEP.

\end{document}